\documentclass[twocolumn]{aa}
\usepackage{amsmath,graphicx,amssymb}
\newcommand \msun {\mbox{{M}$_{\odot}$}}

\newcommand \degree {\mbox{$^\circ$}}

\begin{document}
   \title{Continuum emission in NGC~1068 and NGC~3147:\\ Indications
   for a turnover in the core spectra\thanks{This paper was based on
   observations with the IRAM Interferometer. IRAM is supported by
   INSU/CNRS (France), MPG (Germany) and IGN (Spain).}}


   \author{M.\ Krips
          \inst{1}
          \and
          A.\ Eckart\inst{1}
	  \and 
	  R.\ Neri\inst{2}
	  \and
	  R.\ Sch\"odel\inst{1}
	  \and
	  S.\ Leon\inst{3}
	  \and
	  D.\ Downes\inst{2}
	  \and 
	  S. Garc\'{\i}a-Burillo\inst{4}
	  \and 
	  F. Combes\inst{5}
          }

   \offprints{M.\ Krips\\ 
              \email{krips@ph1.uni-koeln.de}}

   \institute{I. Physikalisches Institut, Universit\"at zu K\"oln,
              Z\"ulpicher Str. 77, 50937 K\"oln, Germany\\
	      \email{krips@ph1.uni-koeln.de; eckart@ph1.uni-koeln.de;
		rainer@ph1.uni-koeln.de}
         \and
              Institut de Radio Astronomie Millim\'etrique, 
              300 rue de la Piscine, 38406 Saint Martin
              d'H\`eres, France\\
	      \email{neri@iram.fr; downes@iram.fr}
         \and
              Instituto de Astrof\'{\i}sica de Andaluc\'{\i}a (CSIC),
              C/ Camino Bajo de Hu\'etor,24, Apartado 3004, 18080 Granada,
              Spain; \email{stephane@iaa.es}
         \and
              Observatorio Astron\'omico Nacional (OAN)-Observatorio 
	      de Madrid, Alfonso XII, 3, 28014 Madrid, Spain\\
	      \email{s.gburillo@oan.es}
	 \and	
              Observatoire de Paris, LERMA, 61 Av. de l'Observatoire,
              75014 Paris, France\\
	      \email{Francoise.Combes@obspm.fr}
   }
   \date{Received ; accepted }

   \abstract{We present new interferometric observations of the
   continuum emission at mm wavelengths in the Seyfert galaxies
   NGC~1068 and NGC~3147. Three mm continuum peaks are detected in
   NGC~1068, one centered on the core, one associated with the jet and
   the third one with the counter-jet. This is the first significant
   detection of the radio jet and counter-jet at mm wavelengths in
   NGC~1068. While the fluxes of the jet components agree with a steep
   spectral index extrapolated from cm-wavelengths, the core fluxes
   indicate a turnover of the inverted cm- into a steep mm-spectrum at
   roughly $\sim$50~GHz which is most likely caused by
   electron-scattered synchrotron emission. As in NGC 1068, the
   spectrum of the pointlike continuum source in NGC~3147 also shows a
   turnover between cm and mm-wavelengths at $\sim$25~GHz resulting
   from synchrotron self-absorption different to NGC~1068.  This
   strongly resembles the spectrum of Sgr~A$^\star$, the weakly active
   nucleus of our own galaxy, and M81$^\star$, a link between
   Sgr~A$^\star$ and Seyfert galaxies in terms of activity sequence,
   which may display a similar turnover.  }

\authorrunning{Krips et al.}
\titlerunning{Continuum emission in NGC~1068 and NGC~3147}
\maketitle
%

\begin{figure*}[!]
\centering
\resizebox{\hsize}{!}{ \rotatebox{-90}{\includegraphics{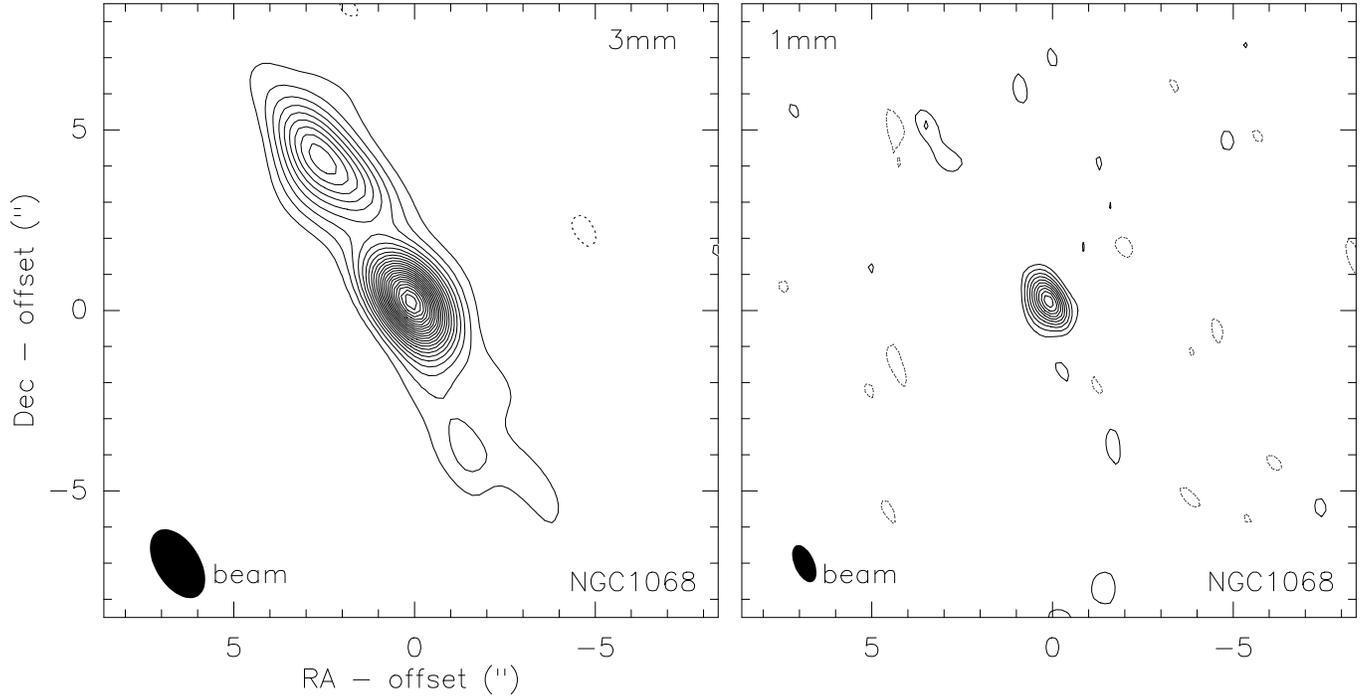}}}
 \caption{Continuum maps of NGC~1068 at 3mm ({\it left panel}) and
     1~mm ({\it right panel}). Contour levels are from $3\sigma$=1.1
     to 25.3~mJy/beam in steps of 3$\sigma$ at 3~mm and from
     $3\sigma$=2.1~mJy/beam to 14.7~mJy/beam in steps of 2$\sigma$ at
     1~mm. The beam size at both wavelengths is indicated as filled
     black ellipse in the lower left corner of the maps: 3mm -
     2.1$''$$\times$1.2$''$, PA=32$^\circ$; 1mm -
     1.0$''$$\times$0.6$''$, PA=36$^\circ$. The (0,0) position of the
     map corresponds to the radio core position (see caption of
     Table~\ref{table1}).}
 \label{n1068-cont}
\end{figure*}

\begin{figure*}[!]
\centering
\resizebox{\hsize}{!}{ \rotatebox{-90}{\includegraphics{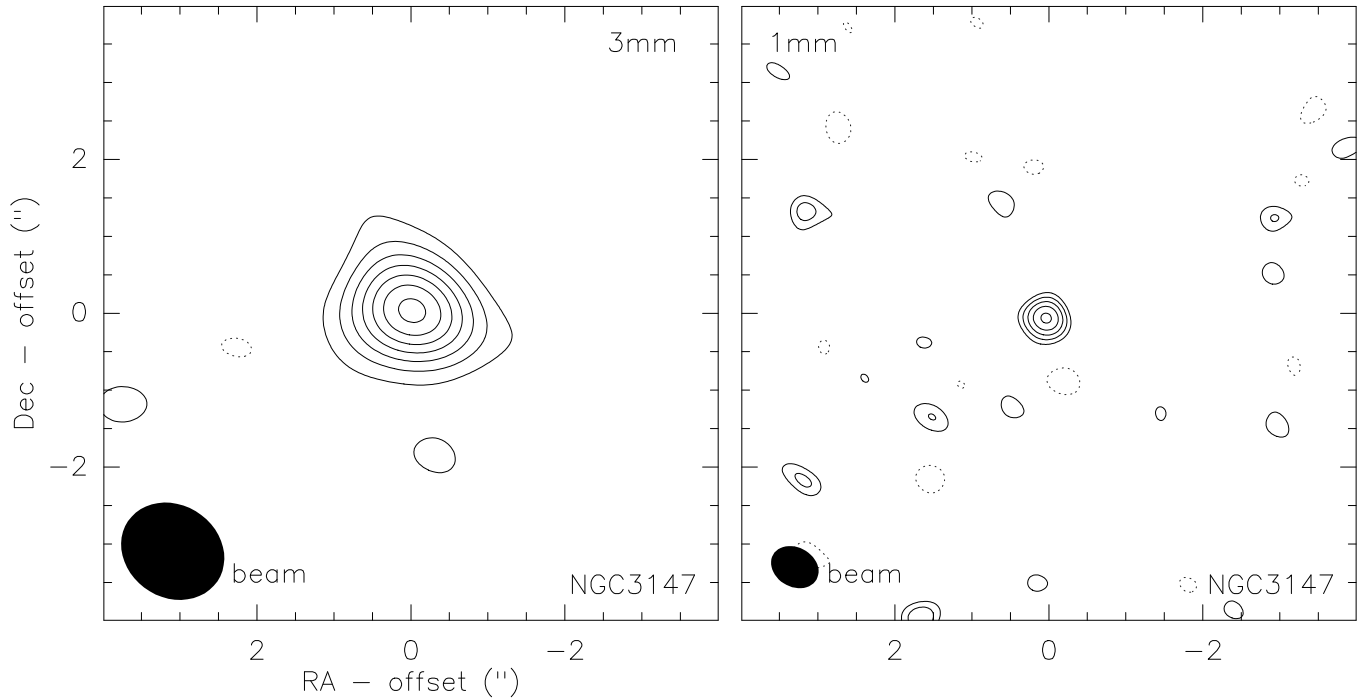}}}
 \caption{Continuum maps of NGC~3147 at 3mm ({\it left panel}) and
     1~mm ({\it right panel}). Contour levels are from $3\sigma$=0.72
     to 4.8~mJy/beam in steps of 3$\sigma$ at 3~mm and from
     $3\sigma$=1.4 to 3.5~mJy/beam in steps of 1$\sigma$ at 1mm. The
     beam size at both wavelengths is indicated as filled black
     ellipse in the lower left corner of the maps: 3mm -
     1.4$''$$\times$1.2$''$, PA=57$^\circ$; 1mm -
     0.7$''$$\times$0.5$''$, PA=60$^\circ$. The (0,0) position of the
     map corresponds to the radio core position (see caption of
     Table~\ref{table1}).}
 \label{n3147-cont}
\end{figure*}

\begin{table*}[!]
\centering
\caption{Continuum parameters for the 3~mm and 1~mm continuum in
NGC~1068 and NGC~3147. Parameters were determined by fitting Gaussian
profiles to the respective components.  }
\begin{tabular}{ccccc}
\hline
\hline
\multicolumn{5}{c}{NGC~1068}\\
\hline
Component & RA & Dec & flux  & Size \\
          & (offset)$^a$ & (offset)$^\mathrm{a}$ & density$^\mathrm{d}$ & 
maj.$\times$min., PA\\
   &  [$''$]  &  [$''$]  & [mJy] & [$''\times''$,\degree]\\
\hline
core-3mm       &  0.1$\pm$0.1 & 0.2$\pm$0.1 & 36$\pm$0.4
               & (0.8$\pm$0.1)\,$\times$\,(0.6$\pm$0.1), 70$\pm$10 \\
jet-3mm        &  2.5$\pm$0.3 & 4.1$\pm$0.3 & 17$\pm$0.5
               & (2.6$\pm$0.1)\,$\times$\,(1.2$\pm$0.1), 39$\pm$2 \\
counterjet-3mm$^{\mathrm{b}}$ 
               & $-$1.4$\pm$0.5   & $-$3.8$\pm$0.5  & 7$\pm$0.6
               &  - \\
core-1mm       &  0.1$\pm$0.2 & 0.2$\pm$0.2 & 22$\pm$0.8
               & (0.6$\pm$0.1)\,$\times$\,(0.47$\pm$0.04), 40$\pm$20 \\
jet-1mm$^{\mathrm{b}}$  &  $\sim$3 & $\sim$5 & 6$\pm$1.0
               & - \\
& \\
\hline
\hline
\multicolumn{5}{c}{NGC~3147}\\
\hline
core-3mm  & 0.0$\pm$0.1 & 0.1$\pm$0.1 & 5.3$\pm$0.2 &
0.1$\pm$0.5$^c$\\
core-1mm  & 0.1$\pm$0.2 & $-$0.2$\pm$0.2  & 2.8$\pm$0.5   & 
0.1$\pm$0.3$^c$\\
\hline
\end{tabular}
\label{contpar}
\begin{list}{}{}
\item[$^{\mathrm{a}}$] offsets with respect to $\alpha_{\rm
  J2000}$=02$^{\rm h}$42$^{\rm m}$40.71$^{\rm s}$ and $\delta_{\rm
  J2000}$=$-00$\degree00$'$47.94$''$ for NGC~1068 (corresponds to the
  position of the radio core; Muxlow et al.\ 1996) and to $\alpha_{\rm
  J2000}$=10$^{\rm h}$16$^{\rm m}$53.65$^{\rm s}$ and $\delta_{\rm
  J2000}$=73\degree24$'$02.70$''$ for NGC~3147 (e.g.\ Ulvestad \& Ho
  2001; Krips et al., in prep.). \\
\item[$^{\mathrm{b}}$] The SNR is not sufficient to estimate a size.\\
\item[$^c$ ] emission most likely point-like\\
\item[$^d$ ] values have been determined by fitting Gaussian profiles
to the visibilities. The uncertainties do not include systematic
errors from the flux calibration ($\sim$10-20\%).
\end{list}
\label{table1}
\end{table*}

\section{Introduction}
NGC~1068 and NGC~3147 are both part of the NU(clei of)-GA(laxy)
project (Garc\'{\i}a-Burillo et al.\ 2003) that recently succeeded in
mapping the molecular gas emission of $^{12}$CO(1--0) and
$^{12}$CO(2--1) in 12 nearby Seyfert and LINER galaxies with high
angular resolution and high sensitivity using the IRAM Plateau de Bure
Interferometer (PdBI). NUGA primarily aims at analysing the dynamics
and distribution of the gas on the smallest possible scales to
disentangle the accretion mechanisms of the gas onto the nucleus which
are supposed to be strongly correlated with the observed nuclear
activity.  As a by-product of this project, information on the
continuum emission at 1~mm and 3~mm was also gained for NGC~1068 and
NGC~3147.

NGC~1068, at a redshift of $z=0.004$, is one of the best studied
Seyfert galaxies. It is regarded to be the archetype for unification
schemes of Seyferts. Khachikian \& Weedman (1974) first classified
NGC~1068 as Seyfert 2 due to the detection of unpolarized
(direct-path) emission lines with narrow widths originating in the
Narrow-Line-Region (NLR). The NLR appears to be extended to the
North-East with respect to the nucleus (e.g.\ Groves et al.\
2004). The radio jet is found to be located in the same direction
(e.g.\ Gallimore et al.\ 2004).  Emission from the broad line region
(BLR) could also be detected through polarized (scattered-path)
emission lines with broad line widths (Antonucci \& Miller et al.\
1985). This strongly suggests that NGC~1068 contains a hidden Seyfert
1 nucleus supporting the unified theory for Seyfert galaxies.  The
position of the nucleus could be precisely derived by radio continuum
observations carried out by Gallimore et al.\ (2004). They report a
compact component with a flat spectrum lying in a maser disk that is
most likely the central engine.  NGC~1068 is one of the few (i.e.\
$\sim30$) extragalactic systems in which maser emission could be
detected. With the aid of H$_2$O maser emission, Greenhill \& Gwinn
(1997) determined the mass of the AGN to 1.7$\times10^7$~\msun in the
central 2.2~pc. The bolometric luminosity of NGC~1068 can be derived
to a few $\sim$10$^{-3}$L$_{\rm edd}$ indicating - in terms of
Eddington luminosity - a slightly higher radiation efficiency of the
accretion and/or a higher accretion rate than in other LLAGN. The
molecular gas in NGC~1068 was mapped by Schinnerer et al.\ (2000, and
references therein). It is distributed in a two arm spiral, along a
bar and in a warped nuclear ring with two emission peaks which are
interpreted as orbit crowding points. Schinnerer et al.\ (2000) report
also on a detection of the 3~mm continuum emission in NGC~1068: two
components become visible at this wavelength, a 16$\sigma$ peak at the
radio core position and a 3$\sigma$ peak at the position of the radio
jet.  Going to longer wavelengths, more continuum structure appears
(e.g., Ulvestad et al.\ 1987, Gallimore et al.\ 1996 \& 2004): the
central continuum peak splits up into four components, of which two
show steep radio spectra (i.e., $f_\nu\propto\nu^\alpha$ with
$\alpha\approx-1$) while the remaining two have flat to inverted
spectra (i.e., $\alpha\approx-0.2$ and $\alpha\approx0.3$). The
inverted spectrum feature (labeled S1 in Gallimore et al.\ 1996) is
associated with the nucleus of NGC~1068 whereas the remaining three
components are supposed to be part of the extended jet and
counter-jet.

NGC~3147, at a redshift of $z=0.009$, is classified as Sbc galaxy
hosting a Seyfert~2 nucleus (Ho et al.\ 1997). ROSAT observations
(Roberts \& Warwick 2000) unveil an X-ray source at the position of
the nucleus which is associated with a supermassive black
hole. Assuming the empirical correlation between black hole mass and
bulge stellar velocity dispersion (e.g., Gebhardt et al.\ 2000,
Ferrarese \& Merritt 2000), the black hole mass can be estimated to
$\sim$4$\cdot10^8$~\msun (Ulvestad \& Ho 2001). The bolometric
luminosity can be determined to $\sim$10$^{-4}$L$_{\rm edd}$ (Ulvestad
\& Ho 2001) indicating a slightly higher radiation efficiency of the
accretion than in other LLAGN similar to NGC~1068. VLBA and VLBI radio
observations also show a pointlike, non-thermal continuum source at
the position of the nucleus (Ulvestad \& Ho 2001, Anderson et al.\
2004, Krips et al.\ in prep.). The radio spectrum is inverted and no
signs of variability are found on timescales of a few years. The CO
emission is astonishingly similar to NGC~1068. Besides a double-peaked
nuclear ring observed in CO(2--1), spiral arms and/or a ring-like
structure become visible at larger scales in IRAM PdBI observations
(Garc\'{\i}a-Burillo et al.\ 2005, Leon et al., in prep.).

\begin{figure}[!]
\centering
 \resizebox{8.7cm}{!}{\rotatebox{-90}{\includegraphics{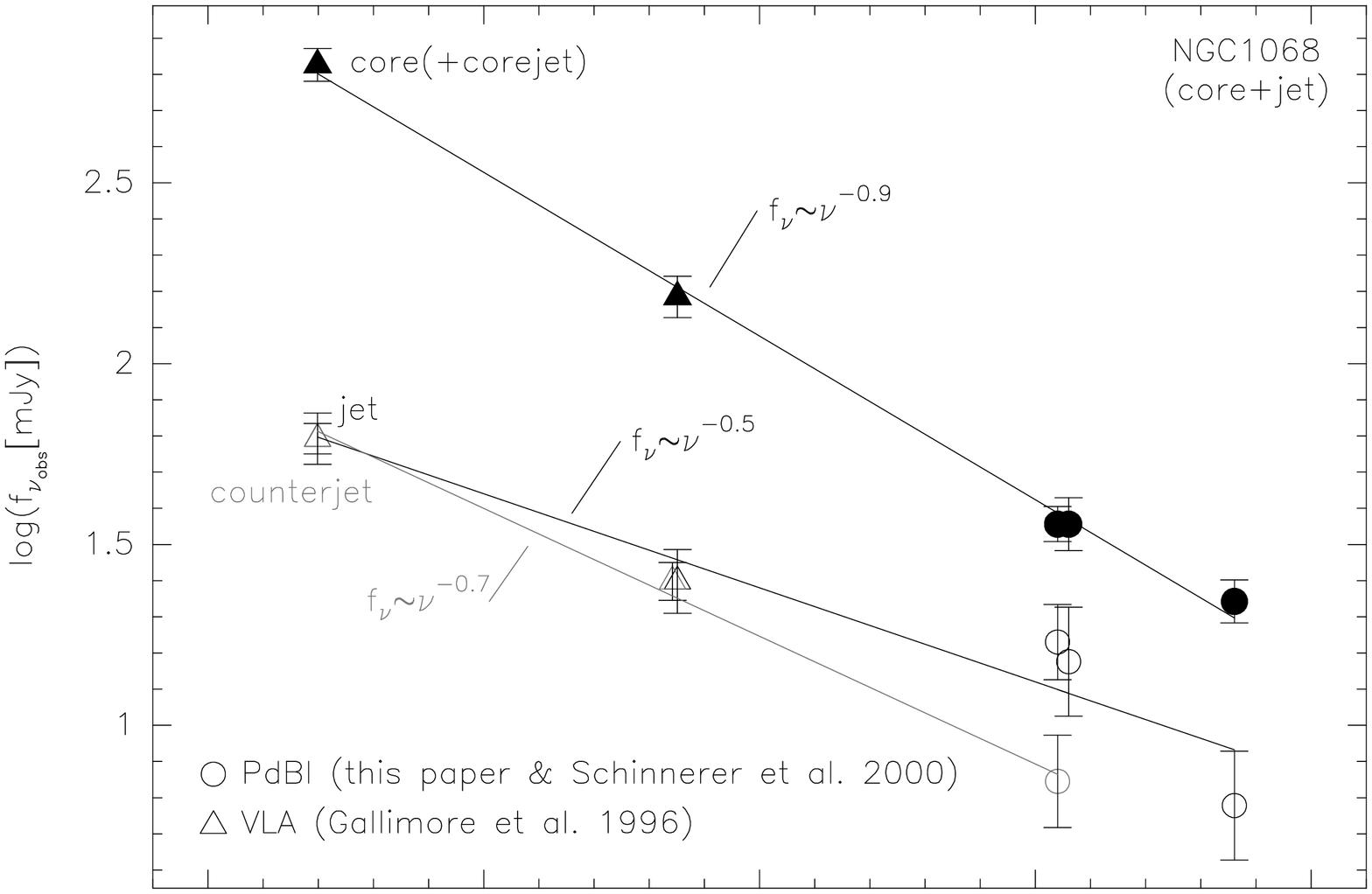}}}
 \resizebox{8.7cm}{!}{\rotatebox{-90}{\includegraphics{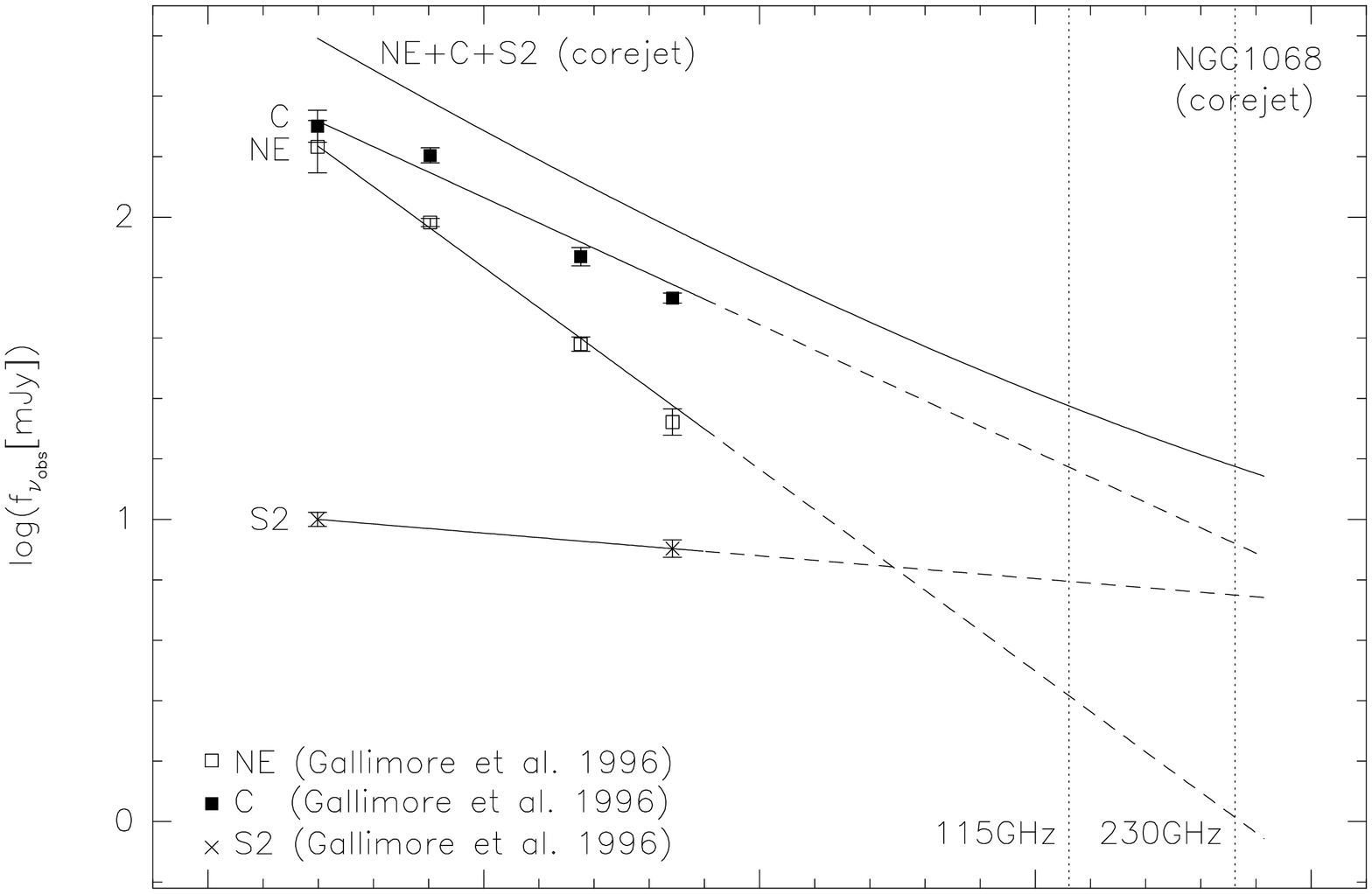}}}
 \resizebox{8.7cm}{!}{\rotatebox{-90}{\includegraphics{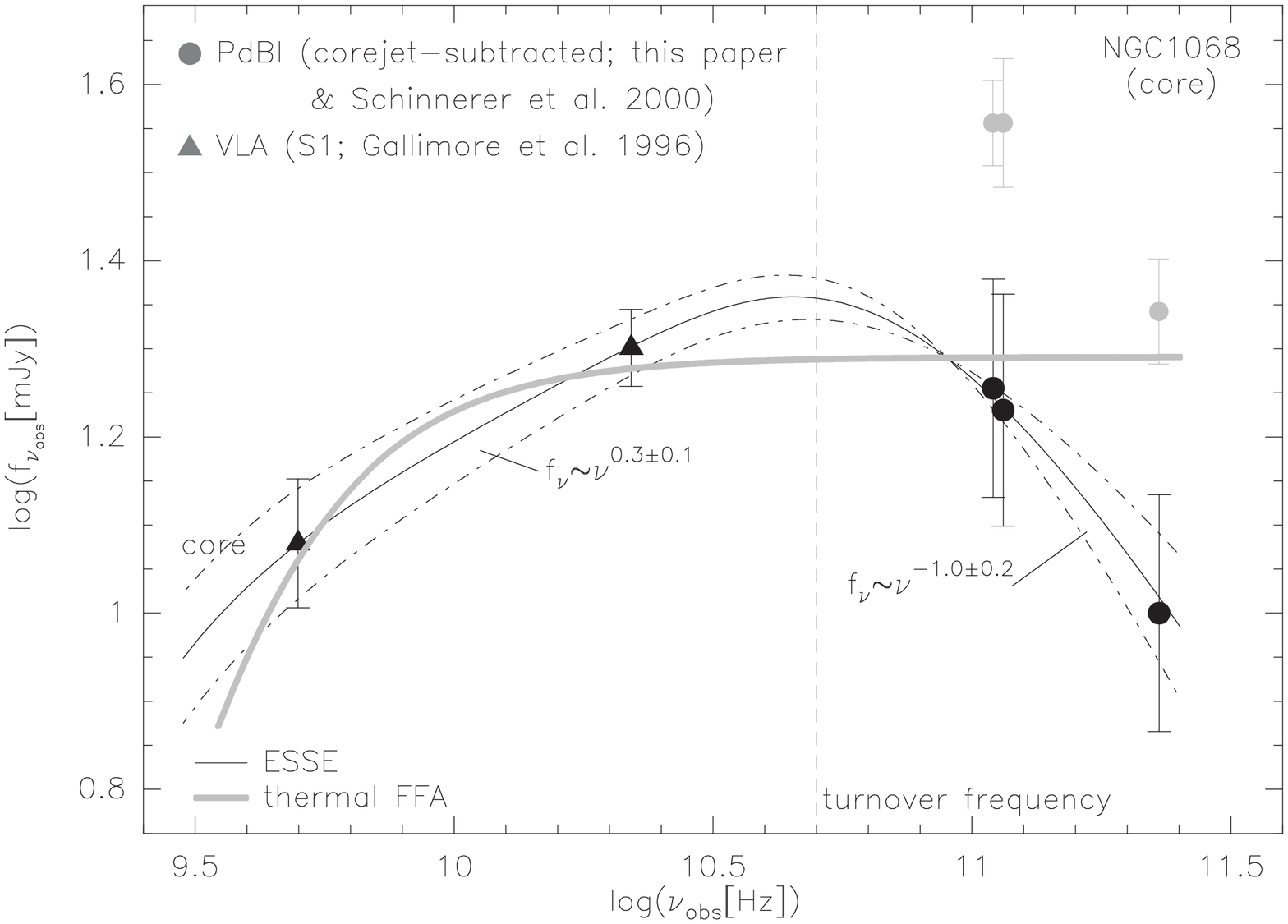}}}
 \caption{SED of the continuum emission in NGC~1068. {\it Upper
 panel:} 230~GHz, 115~GHz and 110~GHz fluxes of the core ({\it filled
 circles}), the jet ({\it open black circles}) and the counter-jet
 ({\it open grey circles}) compared to 22~GHz and 5~GHz VLA
 observations ({\it triangles}; Gallimore et al.\ 1996). The filled
 triangles correspond to fluxes still including contributions from the
 core-jet components visible on VLA scales ($\leq1''$; Gallimore et
 al.\ 1996). {\it Middle panel:} Core-jet components derived by
 Gallimore et al.\ (1996) and extrapolated to mm wavelengths. {\it
 Lower panel:} Pure core spectrum. The filled triangles show the cm
 flux obtained by Gallimore et al.\ (1996; their S1 component). The
 filled circles represent the (estimated) pure core flux at mm
 wavelengths for which the core-jet flux was subtracted (see text for
 more details). The filled grey circles indicate the uncorrected
 mm-fluxes. The dashed-dotted curves (ESSE) show the uncertainties of
 the fit. The thick grey line is the thermal FFA model.}
 \label{n1068-sed}
\end{figure}

\begin{figure}[!]
\centering
\resizebox{\hsize}{!}{\rotatebox{-90}{\includegraphics{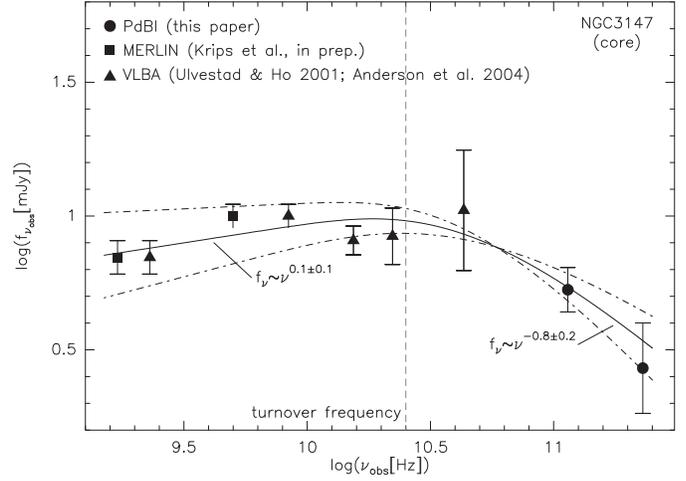}}}
 \caption{Spectral energy distribution of the core emission in
 NGC~3147. The 230~GHz and 110~GHz fluxes of the core (filled circles;
 this paper) are compared with the VLBA 2.3~GHz \& 8.4~GHz flux
 (Ulvestad \& Ho 2001), the VLBA 15.4GHz, 22.2~GHz \& 43.2~GHz flux
 (Anderson et al.\ 2004) and the MERLIN 5~GHz and 1.8~GHz flux (Krips
 et al., in prep.). The dotted-dashed curves indicate the
 uncertainties of the fit.}
 \label{n3147-sed}
\end{figure}

\section{Observations}

\subsection{NGC~1068}
Observations of NGC~1068 at 3~mm and 1~mm were carried out with the
IRAM PdBI in February 2003 using all six antennas in A
configuration. The bandpass was calibrated on NRAO150 while phase and
amplitude calibration were performed on 0235+164 and 0238-084. CRL618
and MWC349 were chosen to determine the absolute flux scale. The
receivers were tuned to the redshifted $^{13}$CO(1--0) line (i.e.,
109.8~GHz) at 3~mm and to the redshifted $^{12}$CO(2--1) line (i.e.,
231.1~GHz) at 1~mm. A total bandwidth of 580~MHz with a spectral
resolution of 1.25~MHz was used per sideband. Applying uniform
weighting\footnote{Actually, when speaking of uniform weighting, it
rather corresponds to robust weighting which is a compromise between
natural and pure uniform weighting. This is a convention in the IRAM
GILDAS Packages.} in the mapping process, beamsizes are derived to
2.1$\times$1.2$''$ with a position angle (PA) of 32\degree at 3mm and
to 1.0$\times$0.6$''$ with PA=36\degree at 1mm.

\subsection{NGC~3147}
$^{12}$CO(1--0) and $^{12}$CO(2--1) emission from NGC~3147 were
observed with the IRAM PdBI in February 2002 and 2004 using all six
antennas in A, B, C, and D configuration. 3C345 was used as bandpass
calibrator, while 1546+027, 1044+719 and 0838+710 were chosen as
amplitude and phase calibrators. The flux was calibrated on MWC349 and
3C273.  The receivers were tuned on the redshifted $^{12}$CO(1--0)
line (i.e., 114.2~GHz) at 3~mm and to the redshifted $^{12}$CO(2--1)
line (i.e., 229.9~GHz) at 1~mm. The total bandwidth was set to 580~MHz
in each sideband with a spectral resolution of 1.25~MHz. The natural
beamsizes are determined to 1.4$''\times$1.2$''$ with PA=57\degree at
3~mm and to 0.7$''\times$0.5$''$ with PA=60\degree at 1~mm.

\section{Results}
The continuum maps at 3~mm and 1~mm were computed from those channels
in which no line emission is expected (i.e., Upper Side Band (USB) at
3~mm and Lower Side Band (LSB) at 1~mm). Continuum emission was
detected at 3~mm and at 1~mm in both sources.

All cm and mm fluxes are accurate at least to within $\sim10-20$\%
accounting also for uncertainties in the flux calibration and
systematic errors.

\subsection{NGC~1068}
\subsubsection{3~mm}
The 3~mm continuum map is shown in Fig.~\ref{n1068-cont} revealing
three different components, one associated with the nucleus, one with
the jet and one with the counter-jet. The jet component is elongated in
North-East direction, agreeing with results found at cm
wavelengths. Also a 6$\sigma$ feature is detected on the opposite side
of the nucleus indicating the counter-jet. Schinnerer et al.\ (2000)
detected the core with a flux of 36$\pm$5~mJy and the jet with
3$\sigma$ to 15$\pm$5~mJy at a slightly higher frequency (115~GHz) but
missed the counter jet. This agrees well in flux and position with our
new data (see Table~\ref{contpar}).

\begin{figure}[!]
\centering
\resizebox{\hsize}{!}{\rotatebox{-90}{\includegraphics{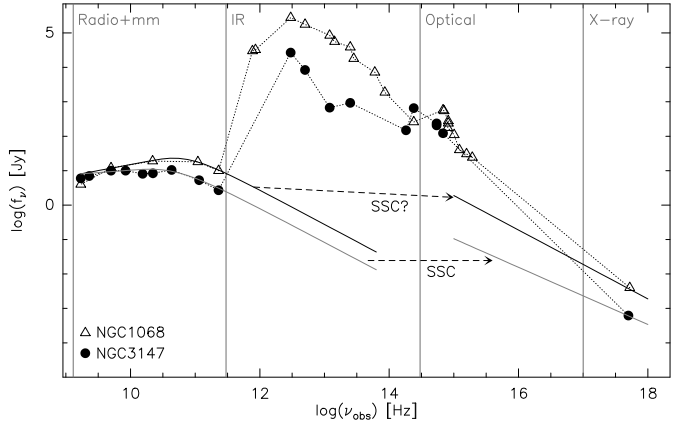}}}
 \caption{Spectral Energy Distribution of NGC~1068 and NGC~3147 from
 radio wavelengths to X-rays. Radio and mm fluxes are from this paper
 while the remaining values are taken from NED and Ulvestad \& Ho
 2001.}
 \label{seda}
\end{figure}

Three elliptical Gaussian profiles were fitted to our 3~mm data. The
results are listed in Table~\ref{contpar}. The position of the core is
determined at $\alpha_{\rm J2000}$=02$^{\rm h}$42$^{\rm m}$40.71$^{\rm
s}$ and $\delta_{\rm J2000}$=$-00$\degree00$'$47.7$''$ consistent with
the core position derived at 6~cm (Gallimore et al.\ 2004).  The jet
and counter-jet fluxes are plotted in Fig.~\ref{n1068-sed} at
different frequencies including radio fluxes estimated by Gallimore et
al.\ (1996). The diagram shows a steep spectral index of
$\alpha\approx-0.5$ ($f_\nu\propto\nu^\alpha$) for the jet and for the
counter-jet. This is consistent with optically thin synchrotron
radiation that is expected for jet emission. The situation for the
core is more complex since VLA maps (Gallimore et al.\ 1996) unveil
further subarcsecond components which cannot be resolved with the IRAM
PdBI: The PdBI ``core'' is resolved by VLA into two further jet
features (NE,C) with steep spectra and two flat to inverted components
(S1,S2) one of which is associated with the nucleus (labeled S1 in
Gallimore et al.\ 1996). An estimate from lower resolution radio maps
(Gallimore et al.\ 1996), not distinguishing between the respective
subarcsecond components, indicates a steep spectrum for the overall
(inner $\sim$2'') region we designated as ``core(+corejet)'' (i.e.\
S1+S2+NE+C; $\alpha$=$-$0.9; upper panel of Fig.~\ref{n1068-sed}).
Thus, optically thin synchrotron radiation from the subarcsecond jet
($\equiv$core-jet; NE+C+S2) most likely dominates the emission at
mm-wavelengths.  However, since the fluxes of the subarcsecond radio
(22~GHz) spots (namely NE,C and S2 in Gallimore et al.\ 1996) are
known, their contribution to our ``core(+corejet)'' mm-component can
be estimated to separate it from the nuclear component S1 and at least
a rough estimate for S1 at 3~mm can be given. Assuming the spectrum
calculated from Gallimore et al.\ (1996; middle panel in
Fig.~\ref{n1068-sed}) for these features, we find that $\sim$60\% of
the 3~mm core(+corejet) still comes from these components (middle
panel of Fig.~\ref{n1068-sed}). Subtracting this value from the 3~mm
fluxes (115 and 110~GHz) allows to compare both with the core (S1)
fluxes at 5~GHz of 12$\pm$1~mJy and at 22~GHz of 19.1$\pm$0.6~mJy
(VLA, Gallimore et al.\ 1996). The lower panel of Fig.~\ref{n1068-sed}
indicates a drop of the flux from the power law spectrum with
$\alpha=0.3$ towards the 3~mm data.  Accounting for the narrower beam
at the cm-wavelengths compared to the mm-observations and thus
resolution effects, the drop might be even more extreme\footnote{Note
that the beamsize at 5~GHz is however larger than at 22~GHz.}. Another
caveat to keep in mind is possible variability.  However, the mm
fluxes give no hints for any variability between different epochs
although a flux variation of $\sim$30\% has been reported for one of
the core-jet components (C) (Gallimore et al.\ 2004). This needs to be
verified with future observations.

\subsubsection{1~mm}
\label{n1068-1mm}
In our PdBI data from 2003, we clearly detect
($>$20$\sigma$$\equiv$22~mJy/beam) continuum emission at 1~mm
(Fig.~\ref{n1068-cont}), while Schinnerer et al.\ (2000) report only
an upper limit of the 1~mm continuum of 6~mJy. However, a refined
analysis of the earlier Schinnerer et al.\ (2000) maps shows that also
their data contain continuum emission which is consistent to 10\% with
what we calculate from the new observations.  The position of the core
derived from the 1~mm data is consistent with the position derived at
3~mm. The flux of the 1~mm core component follows the power law fitted
to the lower frequencies in Fig.~\ref{n1068-sed} (upper panel). One
has to keep in mind, however, that the core emission at 1~mm might
also still be biased by the different core-jet components (NE,C and
S2; Gallimore et al.\ 1996) within the central 1$''$ detected with VLA
and VLBA. Assuming the spectral indices estimated from the radio
fluxes and taking the (VLA) fluxes at 22~GHz (Gallimore et al.\ 1996),
the contribution of these further components (NE,C, S2) to the pure
core (S1) emission at 1~mm can be estimated to $\sim$60\% (middle
panel of Fig.~\ref{n1068-sed}). The flux of the estimated mm core-jet
might be even higher accounting for the higher angular resolution of
the cm data compared to the mm data and thus of resolution effects, so
that the 1~mm flux of the pure core could still be lower. Subtracting
the contribution of the core-jet from the 1mm flux and comparing it
with the core flux derived at cm wavelengths by Gallimore et al.\
(1996), we find further support for a turnover between cm and mm
wavelengths (lower panel of Fig.~\ref{n1068-sed}). Even if we do not
apply a core-jet correction and assume hence the 1~mm emission to be
mainly dominated by the core alone, the 1~mm flux is already
significantly below the inverted power law spectrum produced by the cm
and 3~mm fluxes.  Although resolution effects between 3~mm and 1~mm
have to be taken into account, they should only play a significant
role for the extended core-jet components but not for the core itself
since it appears to be quite compact at cm wavelengths (VLA
scales). Thus, accounting for all arguments, we think that the
turnover is a real intrinsic property of the NGC~1068 nucleus and no
artefact of resolution effects or calibration uncertainties.

In general, three mechanisms can be proposed to explain such a
low-frequency turnover in NGC~1068 (e.g., Roy et al., 1998, Gallimore
et al., 1997 \& 2004): (1) synchrotron emission from a plasma with
electron self-absorption (SSA), (2) synchrotron emission which is
electron-scattered in the obscured central region (ESSE), or (3)
thermal free-free absorption from an X-ray heated (thermal FFA),
ionized plasma. The spectra in all three cases can be fitted with the
following equation (e.g., Kraus 1986):
\begin{eqnarray}
f_\nu & \propto & \left(\frac{\nu}{\nu_0}\right)^{\alpha_{\rm thick}}
        \left(1-\exp{\left[-\left(\frac{\nu_0}{\nu}\right)^{\alpha_{\rm
        thick}-\alpha_{\rm thin}}\right]}\right)\cdot f_\tau,
\label{synch}
\end{eqnarray}
where $\nu_0$ is in units of $10^9$~Hz, and $\alpha_{\rm thick}$ and
$\alpha_{\rm thin}$ are the respective spectral indices for the
optically thick/thin part of the spectrum. In the first case of
synchrotron self absorption (SSA) and also in the third case of
thermal free-free absorption (FFA), $f_\tau$\footnote{$f_\tau$ is only
an artificial parameter to summarize all three cases in one equation.}
is 1. In the second case of electron-scattered synchrotron emission
(i.e.\, free-free absorption by a scattering ionised gas; ESSE),
$f_\tau$ is equal to $\exp(-\tau_{ff}(\nu))$ with:
\begin{eqnarray}
\tau_{ff}(\nu) & = & \tau_0\nu^{-2.1}\,\, =\,\, 
0.08235\times T_e^{-1.35}n_e^2\,L\,\nu^{-2.1}
\end{eqnarray}
$T_e$ is the electron temperature in units of Kelvin, $n_e$ is the
mean square electron density in ${\rm cm^{-2}}$, and $L$ is the path
length along the sight line in pc (e.g., Altenhoff et al., 1960;
Mezger \& Henderson 1967).

Gallimore et al.\ (1996) already excluded the first scenario based on
the low brightness temperatures of $\sim$3$\times10^6$~K derived at
5~GHz. They argue that this temperature is too low for synchrotron
self absorption which generally becomes important at much higher
brightness temperatures ($\gtrsim10^8$~K; e.g., Williams 1963,
Kellerman \& Owen 1989). We therefore consider only the two latter
cases in the following of either electron-scattered synchrotron
emission (ESSE) or thermal free-free absorption (FFA) for the S1
component in NGC~1068. As thermal free-free emission is found to have
generally $\alpha_{\rm thick}=2$ and $\alpha_{\rm thin}=0$ (e.g.,
Kraus 1986 (Orion or Rosette nebula); Roy et al., 1998), this does not
appear to be consistent with the measured spectrum in which the data
point at 1~mm lies significantly below the thermal FFA curve (see
boldfaced grey solid curve in the lower panel of
Fig.~\ref{n1068-sed}). However, one important caveat to be considered
are resolution effects which might result in an artificial drop of the
spectrum between 3~mm and 1~mm. This cannot be totally excluded
although it seems likely that this effect has a stronger influence on
the extended jet and core-jet components than on the rather compact
core (S1). Assuming the ESSE case (see solid and dashed-dotted curves
in the lower panel of Fig.~\ref{n1068-sed}), the observed spectrum can
be nicely reproduced. The fitted optical depth $\tau_{ff}$ is
consistent with an absorbing plasma at $T_e=10^6$K, and an emission
measure of $n_e^2L=10^9$cm$^{-6}$pc agreeing with the values used by
Roy et al. \ (1998) and Gallimore et al.\ (2004). The turnover
frequency can be derived to 50$\pm$10~GHz for NGC~1068. The error was
calculated from the fit assuming equal weighting factors for the data
points.

Besides the core emission, a further weak ($\sim$5$\sigma$) continuum
peak is detected at (roughly) the position of the large scale radio
jet. The flux is compatible with the steep jet spectrum
($f_\nu\propto\nu^{-0.6}$) extrapolated from lower frequencies.

\subsection{NGC~3147}
\subsubsection{3~mm}
The 3~mm continuum appears pointlike consistent with radio
observations at different angular resolutions (VLBA: Ulvestad \& Ho
2001, VLBI \& MERLIN: Krips et al., in prep.). While the radio fluxes
span an inverted spectrum with $\alpha\approx0.1-0.4$
($f_\nu\propto\nu^\alpha$), the 3~mm continuum significantly falls
below the spectrum extrapolated to the mm domain from radio
frequencies (Fig.~\ref{n3147-sed}). This cannot be explained by
resolution effects. First, NGC~3147 is pointlike almost on all angular
scales which is supported by identical radio fluxes at one frequency
but derived from different angular resolutions.  Second, the radio
fluxes derived with MERLIN (i.e., beamsizes are $\lesssim$0.2$''$) are
{\it higher} than the mm-flux determined with the IRAM PdBI (beamsize
is $\sim$1$''$). Variability equally appears to be
unlikely. Observations at different epochs, different wavelengths
(18cm, 6cm, 3mm \& 1mm), and with different instruments (IRAM PdBI,
MERLIN \& VLBI) give no hints for any variation above 10\% of the flux
in NGC~3147, neither at cm nor at mm wavelengths (Ulvestad \& Ho 2001,
Krips et al., in prep.).

\subsubsection{1~mm}
As already expected from 3~mm and cm wavelengths, the continuum
emission at 1~mm is pointlike as well. Based on the results of the
radio data, resolution effects can thus be almost discarded. The flux
derived at 1~mm further decreases compared to 3~mm and higher
wavelengths (see Fig.~\ref{n3147-sed}). Thus, the turnover in the
spectrum from cm to mm wavelengths indeed is an intrinsic property of
the active nucleus in NGC~3147. The relatively high brightness
temperature derived at 5~GHz of at least $\sim10^8$~K (e.g., Anderson
et al.\ 2004, Krips et al., in prep.), suggests synchrotron
self-absorption to be responsible for the turnover in NGC~3147,
different to NGC~1068. By fitting Equation~\ref{synch} to the data,
the turnover frequency can be derived to 25$\pm$10~GHz.

\section{Correlation between radio/mm and X-rays}
\label{corr}
Fig.~\ref{seda} presents the overall spectrum of NGC~1068 and NGC~3147
from radio to X-ray frequencies. For completeness, infrared and
optical fluxes are plotted additionally. The cm/mm data points
correspond to those shown in Fig.~\ref{n1068-sed} (bottom) and
Fig.~\ref{n3147-sed}.

Through Synchrotron Self (inverse) Compton scattering (SSC), the
submm-wavelength photons can be up-scattered into the NIR and X-ray
domain. Using the formula given by Marscher (1983) and Gould (1979),
we can constrain the size of the emitting regions with the observed
X-ray fluxes in NGC~1068 (in the case of ESSE) and NGC~3147.  To
reproduce the observed values, a radius of the order of $\sim$100$R_S$
($R_S$ is the Schwarzschild radius) for the emitting region in
NGC~1068 and NGC~3147 are needed, assuming a Lorentz factor
$\gamma_e=(1-v/c)^{-1/2}$ of a few thousand. While this is consistent
with the radio data of NGC~3147, it appears to be too small for
NGC~1068, relative to the VLBI images of Gallimore et al. (2004),
which suggest a size 10 times larger. Although the X-ray source is
assumed to be of the same order, the radio emission originates from an
area which is at least an order of magnitude larger (Gallimore et al.\
2004). This indicates that, at least in the case of NGC~3147, the
underlying physical mechanisms producing the radio and X-ray emission
are correlated with each other through SSC whereas the situation in
NGC~1068 might be more complicated due to the scattering process and
the equation used above might not be appropriate anymore.  The
correlation between X-rays and radio fluxes via SSC was already
reported for quasars (e.g., Roberts \& Warwick 2000), the equivalents
to NGC~1068 and NGC~3147 at a much higher activity level, and was
recently also found for low-power accreting black holes (e.g., Falcke
et al.\ 2004). As a thermal FFA cannot be totally excluded for
NGC~1068, a correlation between the radio continuum and X-ray emission
is also expected in this case since the absorbing ionized plasma
producing the thermal free-free emission is heated by X-rays coming
from the AGN.

The turnover frequency of a synchrotron spectrum depends on the
magnetic field strength among others. Taking again the equation used
by Gould (1979) and Marscher (1983), the magnetic field density can be
estimated in the case of NGC~3147. However, this estimate also
involves an assumption of the source size and is very sensitive to the
assumed value so that it must be taken with caution and rather
understood as a consistency check. However, assuming SSC mechanism as
discussed above, the X-ray flux set a constraint on the size of the
emitting region for NGC~3147 yielding a radius around 100$R_S$.  If we
assume this size for NGC 3147, then the turnover frequency of 25~GHz
at a flux of 12~mJy implies a magnetic field strength of $\sim$10
Gauss.  This is $\sim$3 times lower than the magnetic field of Sgr A*,
for an accretion rate below the Bondi value (Quataert \& Gruzinov
2000, Eckart et al.\ 2004).

\section{Summary \& Discussion}
Continuum emission was detected in NGC~1068 and NGC~3147 with the IRAM
PdBI at 3~mm and 1~mm. While the maps of NGC~1068 unveil several
extended components in form of a jet, a counter-jet and an extended
core, the continuum emission in NGC~3147 appears pointlike. The jet
and counter-jet in NGC~1068 were detected for the first time with a
large signal-to-noise ratio at mm-wavelengths. Their positions and
fluxes are consistent with values obtained from radio observations.
Extrapolating the influence of the core-jet ($\leq$1$''$) components
seen in VLA maps (Gallimore et al.\ 1996), i.e.\ at higher angular
resolution, on the measured mm-flux at the position of the core, a
turnover in the pure core spectrum is indicated at roughly
$\sim$50~GHz. However, further observations at higher angular
resolution are needed to separate the different components within
1$''$ at mm-wavelengths. This will allow to strengthen the case for
the turnover indicated here. The situation in NGC~3147 is
significantly less speculative. The pointlike structure of its
continuum emission discards the influence of resolution effects on the
derived fluxes induced by different angular resolutions. The spectrum
clearly reveals a turnover from flat (cm) to steep (mm).  The turnover
frequencies can be determined to $\sim$50~GHz for NGC~1068 and to
$\sim$25~GHz for NGC~3147. While the turnover in the spectrum of
NGC~1068 is most likely caused by electron-scattered synchrotron
emission (although thermal free-free absorption cannot be totally
discarded with our new mm-data), the turnover in the spectrum of
NGC~3147 can be associated with synchrotron self-absorption.

The spectra of NGC~1068 and NGC~3147 both remind of the one of
Sgr~A$^\star$ at the center of our own galaxy (e.g.\, Melia \& Falcke
2001).  Also in M81$^\star$, the center of M81, hints have been found
that its spectrum might have a turnover between cm and mm-wavelengths.
The bolometric luminosity (10$^{-5}$L$_{\rm edd}$, Reuter \& Lesch
1996) of M81$^\star$ is intermediate between that of Sgr~A$^\star$
($\sim$10$^{-8}$-10$^{-10}$L$_{\rm edd}$, Melia \& Falcke 2001; Yuan
et al.\ 2003) and the 2 Seyfert galaxies NGC~1068
($\sim$10$^{-3}$L$_{\rm edd}$) and NGC~3147 ($\sim$10$^{-4}$L$_{\rm
edd}$) and can thus be regarded as link between our weak galactic
center and low-luminosity AGN. The similar behaviour of their spectra
might be a further hint for this. Turnovers in the synchrotron spectra
of compact radio sources have been already identified in quasars
before, the other extreme in terms of activity (e.g., Lobanov 1998,
Bloom et al.\ 1999). Giga-Hertz-Peaked (GPS) quasars with turnover
frequencies at a few GHz are known since quite a while but quasar
spectra peaking at several 10~GHz have been found as well (e.g., Bloom
et al., 1999). This further connection from AGN with low radiation
efficiencies of the accretion and/or highly sub-Eddington accretion
rates over those at somewhat higher efficiencies/rates to the most
extreme cases with accretion rates close to the Eddington limit allows
to set tighter constraints on the models suggested to explain the
different activity levels. Sgr~A$^\star$ in particular provides the
unprecedented opportunity to witness the accretion mechanisms at play
on scales which are impossible to reach for extragalactic AGN but
there is hardly any accretion going on in this case. Studies of the
galactic center stimulated the re-emergence of radiatively inefficient
accretion flow models (RIAFs, see Melia \& Falcke 2001; Quataert
2003), including the advection dominated accretion flow approach
(ADAF, e.g., Narayan \& Yi 1994) and its derivatives. These models are
characterized by low efficiency in converting thermal energy into
electromagnetic radiation, but differ in terms of the inner boundary
conditions such as the presence of strong winds (e.g., Blandford \&
Begelman 1999) and/or convection (Quataert \& Gruzinov 1999), each of
which lead to differing predictions of the detailed spectral energy
distribution (SED) and its variability. M81$^\star$ will be an ideal
testbed to broaden the available observational data base and verify,
i.e.\ further constrain the existing models for sub-Eddington
accretion. This will enable to set a closer link to somewhat higher or
more efficiently radiating systems like NGC~1068 and NGC~3147.
Anderson et al.\ (2004) confronted the measured SED of NGC~3147 with a
predicted spectrum from ADAF and jet models. The simple ADAF models
reproduce most of the observed spectrum in NGC~3147 but result in a
too low radio emission. The same is probably true for NGC~1068 when
comparing its spectrum with their models. Including a compact radio
jet to the ADAF model (i.e., the optically thick part of the base of a
(relativistic) jet (see also Athreya et al.\ 1997)), the spectra can
be better simulated. This supports recent studies which have even
found evidence for the jet dominating the nuclear radio emission in
LLAGN over RIAFs (e.g., Nagar et al.\ 2001, Falcke et al.\ 2004) while
emission from black holes accreting close to the Eddington rate are
rather dominated by their accretion disks. NGC~1068 clearly shows jet
emission on different angular scales. Although thermal free-free
emission in the core cannot be totally discarded with our new mm-data,
the mm/cm emission appears to be rather consistent with synchrotron
emission from a compact component in the core (e.g., the base of the
jet) which is electron-scattered by the obscuring torus. NGC~3147 has
so far given no hints for any extended jet down to a scale of
$\sim$1pc but a compact jet cannot be excluded.  Although the black
hole in NGC~3147 is ten times more massive than in NGC~1068, its
bolometric luminosity is an order of magnitude less than in NGC~1068
indicating a lower radiation efficiency of the accretion.

\begin{acknowledgements}
Part of this work was supported by the German
\emph{Son\-der\-for\-schungs\-be\-reich, SFB\/,} project number 494.
SL acknowledges support by DGI Grant AYA 2002-03338 and Junta de
Andaluc\'ia'. We thank the referee, J.S.\ Ulvestad, for useful
comments that helped to improve the discussion of the data. We are
grateful to Linda Tacconi and Eva Schinnerer for providing the reduced
IRAM PdBI data published in S00.

\end{acknowledgements}


\begin{thebibliography}{}
\bibitem{alte67} Altenhoff, W., Mezger, P.G., Wendker, H., \&
  Westerhout, G., 1960, Ver\"off.\ Sternw., Bonn, 59, 48
\bibitem{anto85} Antonucci, R.R.J.\ \& Miller, J.S., 1985, ApJ, 297,
621
\bibitem{bloo99} Bloom, S.D., Marscher, A.P., Moore, E.M., Gear, W.,
Teräsranta, H., Valtaoja, E., Aller, H.D., and Aller, Margo F., 1999,
ApJS, 122, 1
\bibitem{blan99} Blandford, R.D., \& Begelman, M.C., 1999, MNRAS, 303, 1 
\bibitem{ecka04} Eckart, A., Baganoff, F. K., Morris, M., Bautz,
M. W., Brandt, W. N., Garmire, G. P., Genzel, R., Ott, T., Ricker,
G. R., Straubmeier, C., Viehmann, T., Sch\"odel, R., Bower, G. C.,
and Goldston, J. E., 2004, A\&A, 427
\bibitem{falc04} Falcke, H., K\"ording, E., Markoff, S., 2004, A\&A,
  414, 895
\bibitem{ferr00} Ferrarese, L.\ \& Merritt, D., 2000, ApJ, 539, 9
\bibitem{gall96} Gallimore, J.F., Baum, S.A., and O'Dea Ch.P., 1996,
ApJ, 458, 136
\bibitem{gall97} Gallimore, J. F., Baum, S. A., and O'Dea, C. P.,
  1997, Nature, 388, 852
\bibitem{gall04} Gallimore, J.F., Baum, S.A., and O'Dea Ch.P., 2004,
ApJ, 613, 794
\bibitem{gar03} Garc\'{i}a-Burillo, S., Combes, F., Hunt, L. K.,
Boone, F., Baker, A.J., Tacconi, L.J., Eckart, A., Neri, R., Leon,
S., Schinnerer, E., Englmaier, P., 2003, A\&A, 407, 485
\bibitem{grac05} Garc\'{i}a-Burillo, S., Combes, F., Eckart, A.,
Tacconi, L., Hunt, L., Leon, S., Baker, A., Englmaier, P., Boone, F.,
Schinnerer, E., Neri, R., Krips, M., to appear in the proceedings of
``The Evolution of Starbursts'', August 16-20, 2004, Bad Honef,
Germany, ed. Huettemeister, Manthey, Aalto, Bomans
\bibitem{gebh00} Gebhardt, K., Bender, R., Bower, G., Dressler, A.,
Faber, S.M.; Filippenko, A.V.; Green, R., Grillmair, C., Ho, L.C.;
Kormendy, J., Lauer, T.R., Magorrian, J., Pinkney, J., Richstone, D.,
and Tremaine, S., 2000, ApJ, 539, 13
\bibitem{goul79} Gould, R.J.,1979, A\&A, 76, 306
\bibitem{gren97} Greenhill, L.J., Gwinn, C.R., 1997, Ap\&SS, 248, 261
\bibitem{grov04} Groves, B.A., Cecil, G., Ferruit, P., Dopita, M.A.,
2004, ApJ, 611, 786
\bibitem{hoxx97} Ho, L.C., Filippenko, Al.V., Sargent, W.L.W., 1997, 
ApJS, 112, 315
\bibitem{khac79} Khachikian, E.Y., Weedman, D.W., 1974, ApJ, 192, 581
\bibitem{kell89} Kellerman, K.I., \& Owen, F.N., Radio galaxies and
quasars, In Galactic and Extragalactic Radio Astronomy (eds Verschuur,
G.L., \& Kellermann, K.I:), 563-600 (Springer, New York, 1989)
\bibitem{krau86} Kraus, J.D., Radio Astronomy, Cygnus-Quasar Books,
1986
\bibitem{loba98} Lobanov, A.P., 1998, A\&ASS, 132, 261
\bibitem{mars83} Marscher, A.P., 1983, ApJ, 264, 296
\bibitem{meli01} Melia, F.\ \& Falcke, H., 2001, ARA\&A, 39, 309
\bibitem{mezg67} Mezger, P.G., \& Henderson, A.P., 1967, ApJ, 147, 471
\bibitem{muxl96} Muxlow, T.W.B., Pedlar, A., Holloway, A.J.,
  Gallimore, J.F., \& Antonucci, R.R.J., 1996, MNRAS, 278, 854
\bibitem{naga01} Nagar, N.M., Wilson, A.S., \& Falcke, H.,  2001, ApJ,
  559, 87
\bibitem{nara94} Narayan, R., \& Yi, I.,  1994, ApJ, 428, 13
\bibitem{quat00} Quataert, E.\ \& Gruzinov, A., 2000, ApJ, 545, 842
\bibitem{schi00} Schinnerer, E., Eckart, A., Tacconi, L.J., Genzel,
  R.\ \& Downes, D., 2000, ApJ 533, 850 
\bibitem{reut96} Reuter, H.-P.\ \& Lesch, 1996, A\&A, 310, L5
\bibitem{robe00} Roberts, T.P. \& Warwick, R.S., 2000, MNRAS, 315, 98
\bibitem{roya98} Roy, A., Colbert, E.J.M., Wilson, A.S., and Ulvestad,
  J.S., 1998, ApJ, 504, 147
\bibitem{ulve87} Ulvestad, J.S., Neff, S.G., Wilson, A.S., 1987, AJ,
  93, 22
\bibitem{ulve01} Ulvestad, J.S.\ \& Ho, Luis C., 2001, ApJ 562, L133
\bibitem{will63} Williams, P. J. S. 1963, Nature, 200, 56
\bibitem{yuan03} Yuan, F., Quataert, E., and Narayan, R., 2003, ApJ, 
598, 301


\end{thebibliography}
\end{document}